\documentstyle[12pt]{article}
\catcode`\@=11

\@addtoreset{equation}{section}

\def\@normalsize{\@setsize\normalsize{15pt}\xiipt\@xiipt
\abovedisplayskip 14pt plus3pt minus3pt%
\belowdisplayskip \abovedisplayskip
\abovedisplayshortskip  \z@ plus3pt%
\belowdisplayshortskip  7pt plus3.5pt minus0pt}

\def\small{\@setsize\small{13.6pt}\xipt\@xipt
\abovedisplayskip 13pt plus3pt minus3pt%
\belowdisplayskip \abovedisplayskip
\abovedisplayshortskip  \z@ plus3pt%
\belowdisplayshortskip  7pt plus3.5pt minus0pt
\def\@listi{\parsep 4.5pt plus 2pt minus 1pt
            \itemsep \parsep
            \topsep 9pt plus 3pt minus 3pt}}

\def\underline#1{\relax\ifmmode\@@underline#1\else
	$\@@underline{\hbox{#1}}$\relax\fi}
\@twosidetrue

\catcode`@=12
     
\catcode`\@=11

\def\FERMIPUB{}     
\def\FERMILABPub#1{\def\FERMIPUB{#1}}
\def\ps@headings{\def\@oddfoot{}\def\@evenfoot{}
\def\@oddhead{\hbox{}\hfill
	\makebox[.5\textwidth]{\raggedright\ignorespaces --\thepage{}--
	\hfill {\rm FERMILAB--Pub--\FERMIPUB}}}
\def\@evenhead{\@oddhead}
\def\subsectionmark##1{\markboth{##1}{}}
}
     
\ps@headings
     
\catcode`\@=12
     
\relax

\def\figcap{\section*{Figure Captions\markboth
	{FIGURECAPTIONS}{FIGURECAPTIONS}}\list
	{Figure \arabic{enumi}:\hfill}{\settowidth\labelwidth{Figure 999:}
	\leftmargin\labelwidth 
	\advance\leftmargin\labelsep\usecounter{enumi}}}
 \relax
\def\tablecap{\section*{Table Captions\markboth
	{TABLECAPTIONS}{TABLECAPTIONS}}\list
	{Table \arabic{enumi}:\hfill}{\settowidth\labelwidth{Table 999:}
	\leftmargin\labelwidth 
	\advance\leftmargin\labelsep\usecounter{enumi}}}
 \relax
\def\reflist{\section*{References\markboth
	{REFLIST}{REFLIST}}\list
	{[\arabic{enumi}]\hfill}{\settowidth\labelwidth{[999]}
	\leftmargin\labelwidth 
	\advance\leftmargin\labelsep\usecounter{enumi}}}
 \relax

\catcode`\@=11
\def\section{\@startsection{section}{1}{0.0ex}
                   {3.5ex plus -1.0ex minus -0.2ex}
                   {2.3ex plus 0.2ex}{\bf}}
\def\subsection{\@startsection{subsection}{2}{0.0ex}
                        {3.25ex plus 1ex minus .2ex}
                        {1.5ex plus .2ex}{\bf}}
\catcode`@=12
\newskip\humongous \humongous=0pt plus 1000pt minus 1000pt
\def\caja{\mathsurround=0pt}

\newif\ifdtup
\def\panorama{\global\dtuptrue \openup1\jot \caja
	\everycr{\noalign{\ifdtup \global\dtupfalse
	\vskip-\lineskiplimit \vskip\normallineskiplimit
	\else \penalty\interdisplaylinepenalty \fi}}}
\def\eqalignno#1{\panorama \tabskip=\humongous
	\halign to\displaywidth{\hfil$\displaystyle{##}$
	\tabskip=0pt&$\displaystyle{{}##}$\hfil
	\tabskip=\humongous&\llap{$##$}\tabskip=0pt
	\crcr#1\crcr}}

\def\oldreffmt#1{\rlap{[#1]} \hbox to 2\parindent{}}

\def\figfmt#1{\rlap{Figure {#1}} \hbox to 1in{}}

\def\beq{\begin{equation}}
\def\eeq{\end{equation}}

\relax
                  
\begin{document}
\def\thefootnote{\fnsymbol{footnote}}
\FERMILABPub{95/236--T}
\begin{titlepage}
\vspace*{-0.7in}
\begin{flushright}
        FERMILAB--PUB--95/236--T\\
        OSU Preprint 306\\
        July 1995\\
\end{flushright}
\begin{center}
{\bf Construction of an {\mbox{$SO(10) \times U(1)_F$}} Model
	of the Yukawa Interactions}\\
\vskip 0.20in
        {\bf Carl H. ALBRIGHT}\\
 Department of Physics, Northern Illinois University, DeKalb, Illinois 
60115\footnote{Permanent address}\\[-0.2cm]
        and\\[-0.2cm]
 Fermi National Accelerator Laboratory, P.O. Box 500, Batavia, Illinois 
60510\footnote{Electronic address: ALBRIGHT@FNALV}\\
\vskip 0.1in
        {\bf Satyanarayan NANDI}\\
 Department of Physics, Oklahoma State University, Stillwater, Oklahoma
        74078\footnote{Electronic address: PHYSSNA@OSUCC}\\
\end{center}
\vskip 0.5in
\begin{abstract}
\indent  We construct a supersymmetric $SO(10) \times U(1)_F$ model of the 
Yukawa interactions at the grand unification scale from knowledge of a 
phenomenological set of mass matrices obtained by a previous bottom-up
approach.  The $U(1)_F$ family symmetry determines the textures for the 
Majorana and generic Dirac mass matrices, while the $SO(10)$ symmetry relates 
each particular element of the up, down, neutrino and charged lepton Dirac
matrices.  The dominant second and third family contributions in the Dirac
sector are renormalizable, while the remaining contributions to the Dirac 
mass matrices are of higher order, restricted by the $U(1)_F$
family symmetry to a small set of tree diagrams, and mainly 
complex-symmetric.  The tree diagrams for the Majorana mass matrix are all
non-renormalizable and of progressively higher-order, leading to a nearly
geometrical structure.  Pairs of ${\bf 1, 45, 10}$ and ${\bf 126}$ Higgs
representations enter with those having large vacuum expectation values 
breaking the symmetry down to $SU(3)_c \times SU(2)_L \times U(1)_Y$ near the 
grand unification scale.  In terms of 12 parameters expressed as the 
Yukawa couplings times vacuum expectation values for the Higgs representations
employed, a realistic set of 15 quark and lepton  masses (including those for
the 3 heavy righthanded Majorana neutrinos)and 8 mixing
parameters emerges for the neutrino scenario involving the non-adiabatic
conversion of solar neutrinos and the depletion of atmospheric muon-neutrinos
through oscillations into tau-neutrinos. 
\end{abstract}
\end{titlepage}
\setcounter{page}{2}
\section{INTRODUCTION}
The standard model (SM) of strong and electroweak interactions, while 
providing excellent agreement with experiment todate, is known to be 
woefully inadequate to explain the mass spectrum and mixings of the 
three families of quarks and leptons.  One needs to go beyond the standard
model in order to relate the independent Yukawa couplings to each other.
Of the various possibilities, supersymmetric grand unified theories and 
superstring theories seem to hold the most promise for successfully 
incorporating the Yukawa interactions in a more satisfactory fashion.  
In this paper we shall restrict our attention to supersymmetric $SO(10)$
grand unification, which has been shown [1] to unify the gauge couplings 
successfully at a scale of $\Lambda_{SGUT} \sim 10^{16}$ GeV.

It is a generally held opinion that knowledge of the mass matrices in the 
weak flavor basis can provide insights into the dynamical mass-generating 
mechanism.[2]  This follows from the fact that the mass eigenvalues are
obtained by diagonalization of the mass matrices, while the mixing matrices
in the mass eigenbasis can be constructed from knowledge of the diagonalizing
matrices connecting the two bases.  By starting from the correct mass 
matrices, one should then be able to deduce the observed quark and lepton
masses and mixings after evolving the results down to the present ``low
energy'' scales.

Generally two procedures are at one's disposal for the identification of the
``correct'' mass matrices.  One can attempt to postulate a particular 
structure or ``texture'' for the mass matrices based on some well-defined
and presumably simple theoretical concepts such as the unification group
and/or the number of texture zeros present.[3]  This procedure has been 
employed by most researchers in the past twenty years.  Alternatively, one
can make use of the known low energy mass and mixing data, supplemented by
reasonable guesses for data which is not yet well determined, in order 
to extract mass matrices within some framework at the unification scale
which yield the low energy data in question.  Of special interest are 
neutrino scenarios incorporating the Mikheyev - Smirnov - Wolfenstein (MSW) [4]
nonadiabatic resonant conversion interpretation of the depletion of
solar electron-neutrinos [5] and either
the observed depletion of atmospheric muon-neutrinos through oscillations [6]
or neutrinos of satisfactory mass to contribute to the hot component of 
mixed dark matter [7], for example.

In a series of papers [8] the authors have demonstrated the latter ``bottom-up''
approach by making use of Sylvester's theorem [9] to construct mass matrices 
from the low energy masses and mixing matrices evolved to the unification 
scale.  
In doing so, we have attempted to look for simplicity of the mass matrices
in the $SO(10)$ framework while varying the quark and lepton weak bases.
Such simplicity was found for the MSW solar and atmospheric neutrino depletions
in the bases where the up quark and Dirac neutrino
mass matrices are real and diagonal, while the down quark and charged lepton 
matrices are in general complex symmetric.  The right-handed Majorana neutrino
matrix exhibits a simple nearly geometrical texture.

From the phenomenological mass matrices constructed, we have attempted to 
derive mass matrices of similar textures from some well-defined family 
symmetry.  In particular, we find within an $SO(10) \times U(1)_F$ symmetry 
framework that we can reproduce all the known and assumed-known low energy mass
and mixing data for the quarks and leptons in terms of products of Yukawa
couplings and Higgs vacuum expectation values (VEVs).  The $U(1)_F$ symmetry 
controls the textures for the generic Dirac and Majorana mass matrices, 
while $SO(10)$ relates particular elements of the up, down, Dirac neutrino and 
charged lepton mass matrices to each other.   

In this paper we shall present all the details for this model construction 
which were summarized in a short letter submitted elsewhere [10].  Section II
summarizes the bottom-up procedure and the phenomenological mass matrices 
obtained for the neutrino scenario preferred.  The $U(1)_F$ family symmetry
is introduced and applied in the Dimopoulos tree-diagram approach [11] in 
Sect. III for the contributions to the mass matrices.  In Sect. IV the 
diagramatic contributions to the mass matrix elements are explicitly given with 
quantitative results presented in Sect~V.  Our work is summarized in Sect. VI.

\section{PHENOMENOLOGICAL MATRICES from a BOTTOM-UP APPROACH}

We begin by presenting the low scale input and procedure by which we were
able to construct a relatively simple $SO(10)$ set of phenomenological mass 
matrices at the SUSY GUT scale as spelled out in detail in Ref. [8]  The 
relevant framework is assumed to be that of SUSY $SO(10)$ grand unification 
at a scale of $\Lambda_{SGUT} = 1.2 \times 10^{16}$ GeV with supersymmetry
breaking occuring at a scale of 180 GeV, in order that we can use the 
analytical one-loop evolution formulas and results given by Naculich [12].

For the low scale quark data, we assumed the following set of quark masses and
Cabbibo - Kobayashi - Maskawa (CKM) mixing matrix [13]
$$\begin{array}{rlrl}
        m_u(1 {\rm GeV})&= 5.1\ {\rm MeV},& \qquad m_d(1 {\rm GeV})&= 8.9\ 
                {\rm MeV}\nonumber\\
        m_c(m_c)&= 1.27\ {\rm GeV},& \qquad m_s(1 {\rm GeV})&= 175\ {\rm MeV}
                \cr
        m_t(m_t)&= 150\ {\rm GeV},& \qquad m_b(m_b)&\simeq 4.25\ {\rm GeV}\cr
  \end{array}\eqno(2.1a)$$
$$V_{CKM} = \left(\matrix{0.9753 & 0.2210 & 0.0031e^{-i155^o}\cr
                -0.2206 & 0.9744 & 0.043\qquad\cr
                0.011  & -0.041 & 0.999\\[-0.15in]\cr
                \qquad -0.001i &  & \cr}\right) \eqno(2.1b)$$
The light quark masses were chosen to be the central values given by Gasser
and Leutwyler [14], while the heavy physical top mass was set equal to 160 GeV
prior to its discovery yielding a running mass of 150 GeV.  We assumed a 
value of 0.043 for $V_{cb}$, which is now thought to be closer to 0.040, and 
applied strict unitarity to determine 
$V_{ub},\ V_{td}$ and $V_{ts}$.  

The greatest $SO(10)$ simplicity was obtained for the neutrino scenario
incorporating the observed depletion of solar neutrinos [5] through the 
nonadiabatic MSW [4] matter conversion of electron-neutrinos
into muon-neutrinos in the interior of the sun and the depletion of atmospheric
muon-neutrinos through oscillation into tau-neutrinos observed now by several
deep mine collaborations [6].  The central values deduced for these mixing plane
results are 
	$$\begin{array}{rlrl}
	  \delta m^2_{12}&\sim 5 \times 10^{-6}\ {\rm eV^2},&\qquad 
		\sin^2 2\theta_{12}&\sim 0.008\nonumber\\
	  \delta m^2_{23}&\sim 1 \times 10^{-2}\ {\rm eV^2},&\qquad 
		\sin^2 2\theta_{23}&\sim 0.9 \cr \end{array} \eqno(2.2)$$
We took for the lepton input 
$$\begin{array}{rlrl}
        m_{\nu_e}&= 0.5 \times 10^{-6}\ {\rm eV},& \qquad m_e&= 0.511\ {\rm 
                MeV}\nonumber\\
        m_{\nu_{\mu}}&= 0.224 \times 10^{-2}\ {\rm eV},& \qquad m_{\mu}&= 
                105.3\ {\rm MeV}\cr
        m_{\nu_{\tau}}&= 0.105\ {\rm eV},& \qquad m_{\tau}&= 1.777\ {\rm 
                GeV}\cr \end{array}\eqno(2.3a)$$
and 
$$V_{lept} = \left(\matrix{0.9990 & 0.0447 & 0.0076e^{-i155^o}\cr
                -0.0363 & 0.8170 & 0.575\qquad\cr
                0.026   & -0.570 & 0.818\\[-0.15in]\cr
	        \qquad -0.007i & & \cr}\right) \eqno(2.3b)$$

These masses and mixing matrix data were evolved to the SUSY GUT scale
by using formulas given by Naculich [12] as spelled out in detail in Ref. 8.
We could then reconstruct complex-symmetric mass matrices at the SUSY GUT 
scale by making use of Sylvester's theorem [9] as illustrated by Kusenko [15] 
for the quark sector.
The construction is not unique, for one is free to change the quark and 
lepton weak bases by letting two parameters, $x_q$ and $x_\ell$, vary 
independently over their support regions, $0 \leq x \leq 1$.  For $x_q\
(x_\ell) = 0$, the up quark (Dirac neutrino) mass matrix is diagonal; while for
$x_q\ (x_\ell) = 1$, the down quark (charged lepton) mass matrix is diagonal.
One is also free to vary the signs of the mass eigenvalues.

By varying the signs of the mass eigenvalues and the two parameters $x_q$ and
$x_{\ell}$, we then searched for a simple $SO(10)$ structure for the mass
matrices.  The greatest simplicity
occurred with $x_q = 0$ and $x_\ell = 0.93$ corresponding to diagonal 
up quark and Dirac neutrino mass matrices and leading to 
	$$M^U \sim M^{N_{Dirac}} \sim diag(\overline{126};\ \overline{126};
		\ 10)\eqno(2.4a)$$
	$$M^D \sim M^E \sim \left(\matrix{10',\overline{126} & 10',
		\overline{126}' & 10'\cr 10',\overline{126}' & \overline{126} 
		& 10'\cr 10' & 10' & 10\cr}\right)\eqno(2.4b)$$
with $M^D_{11},\ M^E_{12}$ and $M^E_{21}$ anomalously small and only the 
13 and 31 elements complex.    Entries in the matrices stand for the Higgs
representations contributing to those elements, which we elaborate upon in
the next Section.  
We have assumed complete unification for the Yukawa couplings of the third
families of quarks and leptons and that vacuum expectation values (VEVs) 
develop only for the symmetric representations ${\bf 10}$ and ${\bf 126}$.
The ${\bf 10}$'s contribute equally to
$(M^U,\ M^D)$ and $(M^{N_{Dirac}},\ M^E)$, while the ${\bf 126}$'s
weight $(M^U,\ M^D)$ and $(M^{N_{Dirac}},\ M^E)$ in the ratio of 
1\ : -3.  The Majorana neutrino mass matrix $M^R$, 
determined from the seesaw formula [16] with use of $M^{N_{Dirac}}$ and the 
reconstructed light neutrino mass matrix, exhibits a nearly geometrical 
structure [17] given by 
$$M^R \sim \left(\matrix{F & - \sqrt{FE} & \sqrt{FC}\cr
                - \sqrt{FE} & E & -\sqrt{EC}\cr
                \sqrt{FC} & -\sqrt{EC} & C\cr}\right)
                        \eqno(2.4c)$$
where $E \simeq {5\over{6}}\sqrt{FC}$ with all elements relatively real [18].  
It can not be purely geometrical, however, since the singular rank-1 matrix 
can not be inverted as required by the seesaw formula, $M^{N_{eff}} \simeq 
-M^{N_{Dirac}}(M^R)^{-1} M^{N^T_{Dirac}}$.

\section{$U(1)_F$ FAMILY SYMMETRY and RESULTING TREE DIAGRAMS}

The challenge is now to introduce a family symmetry which will enable us
to derive the mass matrix textures derived above phenomenologically from
our bottom-up approach.  For this purpose, we propose to use a $U(1)_F$ family
symmetry [19], where we leave open for the time being whether the symmetry is 
global or local in which case it can be gauged.  Before proceeding with this, 
we 
review briefly the elements of the $SO(10)$ symmetry group which play important
roles in our model construction.

In the $SO(10)$ framework, each family of left-handed quarks, leptons,
conjugate quarks and conjugate leptons belongs to a ${\bf 16}$ dimensional
representation.  It is convenient to represent a given flavor (and color) 
member of the ith family and its conjugate by the two components 
$\Psi_{iL} = (\psi_{iL}, (\psi^c)_{iL})$.  In the corresponding 
three-family basis ordered as follows, 
$\Psi_L = \{\psi_{1L},\psi_{2L},\psi_{3L},
(\psi^c)_{1L},(\psi^c)_{2L},(\psi^c)_{3L}\}$, the contributions to the up
or down quark, neutrino or charged lepton, mass terms in the Yukawa Lagrangian
are then given by 
	$${\cal L} = \Psi^T_L C^{-1}{\cal M}\Psi_L + {\rm h.c.}\eqno(3.1a)$$
where the 6 x 6 matrix can be written in terms of 3 x 3 submatrices
	$$\cal{M} = \cal{M}^T = \left(\matrix{M^L & M_{Dirac}\cr
		M^T_{Dirac} & M^R\cr}\right)\eqno(3.1b)$$
with the individual contributions referring to 
	$$\begin{array}{rl}
	  M^L:& (\psi_{iL})^T C^{-1} \psi_{jL} \nonumber\\
	  M_{Dirac}:& (\psi_{iL})^T C^{-1}(\psi^c)_{jL} = \overline{\psi_{jR}}
		\psi_{iL} \cr
	  M^T_{Dirac}:& {(\psi^c)_{iL}}^T C^{-1} \psi_{jL} =
		\overline{\psi_{iR}}\psi_{jL}\cr
	  M^R:& (\psi^c)_{iL}^T C^{-1}(\psi^c)_{jL} = \overline{\psi_{iR}}
		(\psi^c)_{jL}\cr
	\end{array}\eqno(3.1c)$$
Here the diagonal block entries appear only for neutrinos with $M^L$ the 
left-handed Majorana neutrino mass matrix which we take to vanish, while 
$M^R$ is the right-handed Majorana neutrino mass matrix which receives large 
contributions near the SUSY GUT scale.  

By construction the 6 x 6 matrix ${\cal M}$ is complex symmetric, but the Dirac
mass submatrix is not necessarily complex symmetric.  We shall assume that
the dominant contributions are complex symmetric and that any departures 
from this form arise from small higher-order corrections.
Recall that the $SO(10)$ product rules read
	$$\eqalignno{
	{\bf 16} \times {\bf 16} &= {\bf 10}_s + {\bf 120}_a + {\bf 126}_s
		&(3.2a)\cr
        {\bf 16} \times {\bf \overline{16}} &= {\bf 1} + {\bf 45} + {\bf 210}
		&(3.2b)}$$
Hence we shall assume that only the symmetric Higgs representations ${\bf 10}$
and ${\bf 126}$ develop low scale VEVs, while the antisymmetric ${\bf 120}$
does not.  In terms of the $SU(5)$ decompositions, we have 
	$${\bf 10} \rightarrow 5 + \bar{5},\qquad
	  {\bf 126} \rightarrow \overline{50} + 45 + \overline{15} + 10 + 
		\bar{5} + 1\eqno(3.3a)$$
The up-type quarks and Dirac neutrinos then can receive contributions from 
the neutral members of ${\bf 10}(5)$ and $\overline{\bf 126}(5)$, the 
down-type quarks and charged leptons from those of ${\bf 10}(\bar{5})$ and 
$\overline{\bf 126}(\overline{45})$, and the heavy right-handed 
Majorana neutrinos from those of $\overline{\bf 126}(1)$.  We shall later 
assume the Higgs representations ${\bf 1}$ and ${\bf 45}$ play a role in the 
higher-order corrections, where the ${\bf 45}(1)$ and ${\bf 45}(24)$ develop
VEVs according to the decomposition
	$${\bf 45} \rightarrow 24 + 10 + \overline{10} + 1 \eqno(3.3b)$$

Returning to the phenomenological mass matrices obtained in Section II, we
use the textures given in (2.4a,b,c) as our starting point for the construction
of an $SO(10) \times U(1)_F$ model of the Yukawa interactions.  We find it
useful to introduce a generic Dirac matrix, $M_{Dirac}$, to go along with the 
one Majorana matrix, $M^R$.  The $U(1)_F$ family symmetry will then determine
the textures for $M_{Dirac}$ and $M^R$, while the $SO(10)$ symmetry will
relate the corresponding matrix elements of the four Dirac matrices $M^U,\ 
M^D,\ M^N$ and $M^E$ to each other.

Simplicity of the $SO(10)$ structure requires that just one Higgs ${\bf 10}$
representation contributes to the $(M_{Dirac})_{33}$ element (hereafter 
labeled D33).  Since a ${\bf 10}$ contributes equally to the 33 elements of 
all four Dirac matrices, this implies that we assume complete unification of 
the Yukawa couplings at the unification scale: $\bar{m}_{\tau} = \bar{m}_b = 
\bar{m}_t/\tan \beta_{10}$, where $\tan \beta_{10}$ is equal to the ratio 
of the up quark to the down quark VEVs in the ${\bf 10}$, i.e.,
$$\begin{array}{rl}
	\bar{m}_t&= g_{10}(v/\surd{2})\sin \beta_{10} \equiv g_{10}v_u
	\nonumber\\
	\bar{m}_b = \bar{m}_{\tau}&= g_{10}(v/\surd{2})\cos \beta_{10} 
		\equiv g_{10}v_d \cr
	\tan \beta_{10}&= v_u(5)/v_d(\bar{5})\cr 
	\end{array} \eqno(3.4a)$$
in terms of the $SU(5)$ decomposition of $SO(10)$ with $v = 246$ GeV.
The same ${\bf 10}$ can not contribute to D23 = D32, for the diagonal nature
of $M^U$ and $M^N$ requires the presence of another ${\bf 10'}$ with 
	$$ \tan \beta_{10'} = v'_u(5')/v'_d(\bar{5'}) = 0 \eqno(3.4b)$$
Likewise we assume a pure ${\bf \overline{126}}$ contribution to D22 with 
	$$ \tan \beta_{\overline{126}} = w_u(5)/w_d(\overline{45}) \eqno(3.4c)$$
In other words, for simplicity we have taken the 2-3 sector of $M_{Dirac}$
to be renormalizable with two ${\bf 10}$'s and one $\overline{\bf 126}$
developing low scale VEVs.  We illustrate the renormalizable
3-point tree diagrams in Fig. 1a.

We now assign $U(1)_F$ charges as follows to the three families (in order of 
appearance) and to the three Higgs representations introduced which generate
low scale VEVs with the numerical values to be determined later:
	$${\bf 16}_3^{\alpha},\ {\bf 16}_2^{\beta},\ {\bf 16}_1^{\gamma},
		\ {\bf 10}^a, \ {\bf 10'}^b,\ {\bf \overline{126}^c}
		\eqno(3.5a)$$
Conservation of $U(1)_F$ charges then requires $2\alpha + a = 0,\ \alpha + 
\beta + b = 0$ and $2\beta + c = 0$ as seen from the diagrams in Fig. 1a.

We assume the rest of the $M_{Dirac}$ elements arise from higher-order
tree diagrams as first suggested by Dimopoulos [11] twelve years ago.  The 
point is
that not only does SUSY control the running of the Yukawa couplings between
the SUSY GUT scale and the weak scale where it is assumed to be softly 
broken, but it also allows one to assume that only simple tree diagrammatic
contributions to the mass matrices need be considered as a result of the 
non-renormalization theorem [20] applied to loop diagrams.  While the low-scale
VEVs introduced act only once in each diagram, other GUT scale VEVs arising
from ${\bf 1}$ and ${\bf 45}$ Higgs representations can connect superheavy
GUT scale ${\bf 16}$ fermions and their conjugate $\overline{\bf 16}$ mirrors 
to each other and to the three light ${\bf 16}$ families.  The superheavy
fermions and their mirrors pair off and get masses slightly higher than the 
SUSY GUT breaking scale, so their propagators in the higher-order tree 
diagrams lead to non-renormalizable contributions scaled by their masses.  

For each ${\bf 45}$ Higgs representation, as noted earlier in terms of the 
SU(5) decomposition given in (3.3b), VEVs can develop in the orthogonal 
directions 
	$$<{\bf 45}_X> \sim {\bf 45}(1),\qquad <{\bf 45}_Y> \sim {\bf 45}(24)
		\eqno(3.6a)$$
or in any non-orthogonal directions.  One such direction of interest corresponds
to the hypercharge direction for flipped [21] $SU(5) \times U(1)$ as clarified 
in Table I:
	$$<{\bf 45}_Z> = {6\over{5}}<{\bf 45}_X> - {1\over{5}}<{\bf 45}_Y>
		\eqno(3.6b)$$
While the $<{\bf 45}_X>$ VEV breaks $SO(10) \rightarrow SU(5)$,
the $<{\bf 45}_Z>$ VEV breaks $SO(10) \rightarrow {\rm flipped}\ SU(5)$. 
Alternatively, if the $SO(10) \rightarrow SU(5)$ breaking occurs earlier by 
some other VEV such as $<\overline{126}'>$ as required later for the Majorana
sector, the combined action of $<{\bf 45}_X>$ and $<{\bf 45}_Z>$ will
result in the breaking of $SU(5) \rightarrow SU(3)_c \times SU(2)_L \times 
U(1)_Y$.

Since the D13 and D23 elements in (2.4a,b) have the same ${\bf 10'}$
transformation property, this suggests that we introduce a ${\bf 45}^e_X$ 
Higgs field and construct an explicitly complex-symmetric dimension-6 tree 
diagram as shown in Fig. 1b, for which $U(1)_F$ charge conservation requires
$\alpha + \gamma + b + 2e = 0$.  We shall later give the four Dirac mass
matrix contributions derived from D13 by use of Table I which confirms that
D13 and D23 do have the same ${\bf 10'}$ transformation property, i.e., the 
contributions to $M^U$ and $M^N$ vanish while those to $M^D$ and $M^E$ are
non-zero and equal.  

The D12 element, on the other hand, appears to arise from a linear combination 
of ${\bf 10'}$ and new $\overline{\bf 126}$ VEV contributions for which 
$(M^E)_{12} \ll (M^D)_{12}$.  Rather than introduce another new renormalizable 
diagram, we can make use of the fact that a ${\bf 45}_Z$
Higgs develops a VEV which vanishes for the charged lepton D12 diagram as seen
from Table I.  We then introduce a new ${\bf 45}^h_Z$ Higgs field
and construct the complex-symmetric dimension-6 tree diagram shown in Fig. 1b.
Note that detailed study showed that to reduce the number of contributing
diagrams the ${\bf 10'}$ Higgs line should leave the 
diagram, or equivalently, the ${\bf 10'}^*$ line should enter the diagram, so
$U(1)_F$ charge conservation requires $\beta + \gamma - b + 2h = 0$.

The D11 element is dimension-8 or higher, and we leave it unspecified. The 
complex-symmetric leading-order Yukawa diagrams which we wish to generate are 
then neatly summarized by the ordering of the Higgs fields where all external
lines enter the diagrams:
	$$\begin{array}{rl}
	  D33:&  {\bf 16}_3 \ {\bf - 10 - 16}_3\nonumber \\
	  D23:&  {\bf 16}_2 \ {\bf - 10' - 16}_3\\
	  D32:&  {\bf 16}_3 \ {\bf - 10' - 16}_2\\
	  D22:&  {\bf 16}_2 \ {\bf - \overline{126} - 16}_2\\
	  D13:&  {\bf 16}_1 \ {\bf - 45}_X\ {\bf- 10' - 45}_X\ {\bf - 16}_3\\
	  D31:&  {\bf 16}_3 \ {\bf - 45}_X\ {\bf - 10' - 45}_X\ {\bf - 16}_1\\
	  D12:&  {\bf 16}_1 \ {\bf - 45}_Z\ {\bf - 10'^* - 45}_Z\ {\bf - 16}_2\\
	  D21:&  {\bf 16}_2 \ {\bf - 45}_Z\ {\bf - 10'^* - 45}_Z\ {\bf - 16}_1\\
	  \end{array} \eqno(3.7a)$$                             

In order to obtain a different set of diagrams and hence a different texture
for the Majorana matrix, we begin the M33 contribution with a dimension-6 
diagram shown in 
Fig. 1c by including a new ${\bf \overline{126}'}^d$ Higgs which develops a VEV 
at the GUT scale in the $SU(5)$ singlet direction, along with a pair of
${\bf 1}^g$ Higgs fields.  Here $2\alpha + d + 2g = 0$.  The nearly geometric
structure [8] for $M^R$ can then be generated by appending more
Higgs fields to each diagram.  For M23 we introduce another ${\bf 1'}^f$ Higgs
field to construct a diagram with one ${\bf \overline{126}'}^d$, one 
${\bf 45}_X^e$, one ${\bf 1'}^f$ and two ${\bf 1}^g$ fields with charge
conservation demanding $\alpha + \beta + d + 2g + e + f = 0$.  The new
${\bf 1'}$ field is needed in order to scale properly the Majorana matrix
elements relative to each other.  The remaining leading-order diagrams of the 
complex-symmetric Majorana mass matrix
follow by appending more ${\bf 45}_X^e$, ${\bf 45}_Z^h$ and ${\bf 1'}^f$
Higgs lines.  The pattern is made clear from the charge conservation equations:
$2\beta + d + 2g + 2e + 2f =0$ for M22, $\alpha + \gamma + d + 2g + e + h
+ 2f = 0$ for M13, $\beta + \gamma + d + 2g + 2e + h + 3f = 0$ for M12,
and $2\gamma + d + 2g + 2e + 2h + 4f = 0$ for M11.  

In summary, the following
Higgs representations have been introduced in addition to those in (3.5a):
	$${\bf \overline{126}}'^d,\ {\bf 45}_X^e,\ {\bf 45}_Z^h,\ {\bf 1}^g,
		\ {\bf 1'}^f \eqno(3.5b)$$
all of which generate massive VEVs near the GUT scale.  In order to obtain
CP-violation in the quark and lepton mixing matrices, we allow the VEVs for
${\bf 45}_X,\ {\bf 45}_Z,\ {\bf 1}$ and ${\bf 1'}$ to be complex, but the 
VEVs associated with the ${\bf 10,\ 10'\ , \overline{126}}$ and ${\bf 
\overline{126}'}$ representations can be taken to be real without loss of 
generality as seen from our bottom-up results.
Clearly, many permutations of the Higgs fields are possible in 
the higher-order diagrams.

At this point a computer search was carried out to generate $U(1)_F$ charge
assignments leading to the fewest additional diagrams allowed by charge
conservation.  An especially interesting charge assignment stood out for which 
	$$\begin{array}{rl}
	       &\alpha = 9,\ \beta = -1,\ \gamma = -8 \nonumber \\
	&a = -18,\ b = -8,\ c = 2,\ d = -22,\ e = 3.5,\ f = 6.5,\ g = 2.0,\ 
	  	h = 0.5\cr
	\end{array} \eqno(3.8a)$$
One should note that since $\alpha + \beta + \gamma = 0$, the $\left[SO(10)
\right]^2 \times U(1)_F$ triangle anomaly vanishes, whereas the 
$\left[U(1)_F\right]^3$ anomaly does not.  Simplicity then suggests that the 
$U(1)_F$ family symmetry group can be global with a familon being generated 
upon its breaking.  Alternatively, the $U(1)_F$ group can be local and 
gauged if the $\left[U(1)_F\right]^3$ anomaly is canceled by the addition
of several $SO(10)$-singlet fermions with appropriate $U(1)$ charges, or 
perhaps better still, by the Green-Schwarz mechanism [22] provided the model can
be derived from string theory.  We intend to study this point at greater 
length elsewhere and do not commit ourselves here to either possibility.

With the above charge assignments we can further greatly limit the number of 
permutations and eliminate other unwanted diagrams by restricting the $U(1)_F$
charges appearing on the superheavy internal fermion lines.  With the following
minimum set of allowed charges for the left-handed superheavy fermions $F_L$
and their mirror partners $F^c_L$
	$$\begin{array}{rrrrrrrrrr}
	F_L:& -0.5,& 1.0,& 2.0,& 4.0,& 4.5,& -4.5,& -7.5,& 11.0,& 12.5 \nonumber
		\\
      F^c_L:& 0.5,& -1.0,& -2.0,& -4.0,& -4.5,& 4.5,& 7.5,& -11.0,& -12.5\cr
	\end{array} \eqno(3.8b)$$
as determined by another computer program, we recover just the leading-order 
diagrams listed in (3.7a) for the generic 
Dirac mass matrix together with the following uniquely-ordered diagrams for the 
complex-symmetric Majorana mass matrix
	$$\begin{array}{rl}
	  M33:&  {\bf 16}_3 \ {\bf - 1 - \overline{126}' - 1 - 16}_3\nonumber \\
	  M23:&  {\bf 16}_2 \ {\bf - 1 - 45}_X\ {\bf - 1' - \overline{126}' - 1
		 - 16}_3\\
          M32:&  {\bf 16}_3 \ {\bf - 1 - \overline{126}' - 1' - 45}_X\ {\bf - 1
		 - 16}_2\\
	  M22:&  {\bf 16}_2 \ {\bf - 1 - 45}_X\ {\bf - 1' - \overline{126}' - 
		1' - 45}_X\ {\bf - 1 - 16}_2\\
	  M13:&  {\bf 16}_1 \ {\bf - 45}_X\ {\bf - 1' - 1- 45}_Z\ {\bf - 1' - 
		\overline{126}' - 1 - 16}_3\\
          M31:&  {\bf 16}_3 \ {\bf - 1 - \overline{126}' - 1' - 45}_Z\ {\bf - 
		1 - 1' - 45}_X\ {\bf - 16}_1\\
	  M12:&  {\bf 16}_1 \ {\bf - 45}_X\ {\bf - 1' - 1- 45}_Z\ {\bf - 1' - 
		\overline{126}' - 1' - 45}_X\ {\bf - 1 - 16}_2\\
          M21:&  {\bf 16}_2 \ {\bf - 1 - 45}_X\ {\bf - 1' - \overline{126}' - 
		1' - 45}_Z\ {\bf - 1 - 1' - 45}_X\ {\bf - 16}_1\\
	  M11:&  {\bf 16}_1 \ {\bf - 45}_X\ {\bf - 1' - 1- 45}_Z\ {\bf - 1' - 
		\overline{126}' - 1' - 45}_Z\ {\bf - 1 - 1' - 45}_X 
		{\bf - 16}_1\\
	  \end{array} \eqno(3.7b)$$
Several other higher-order diagrams are allowed by the $U(1)_F$ charges given
in (3.8a,b) and appear for D11, D22, M23 and M32 with the Higgs fields ordered
as follows:
	$$\begin{array}{rl}
	  D11:&  {\bf 16}_1 \ {\bf - 45}_X\ {\bf - 1' - 1 - 10' - 1 
		- 1' - 45}_X\ {\bf - 16}_1\nonumber \\
	  D22:&  {\bf 16}_2 \ {\bf - 45}_Z\ {\bf - 10'^* - 1'^* - 16}_2,\quad 
	      	{\bf 16}_2 \ {\bf - 1'^* - 10'^* - 45}_Z\ {\bf - 16}_2\\
	  M23:&  {\bf 16}_2 \ {\bf - 45}_X^*\ {\bf - 1' - 1 - 45}_Z\ {\bf - 1' 
		- \overline{126}' - 1 - 16}_3\\
          M32:&  {\bf 16}_3 \ {\bf - 1 - \overline{126}' - 1' - 45}_Z\ {\bf - 
		1 - 1' - 45}_X^*\ {\bf - 16}_2\\
	  \end{array} \eqno(3.7c)$$
These corrections to M23 and M32 ensure that $M^R$ is rank 3 and nonsingular,
so that the seesaw formula [16] can be applied.  Up to this point the 
contributions are all complex-symmetric.

Additional correction terms of higher order which need not be complex-symmetric 
can be generated for the Dirac and Majorana matrix elements, if one allows 
additional superheavy fermion pairs with new $U(1)_F$ charges.  Such a subset 
which does not destroy the pattern constructed above, but helps to improve the 
numerical results for the lepton masses and mixings, consists of the following:
	$$\begin{array}{rrrrrr}
	F_L:&  1.5,& -6.0,& -6.5 \nonumber \\
      F^c_L:& -1.5,&  6.0,&  6.5 \cr
	\end{array} \eqno(3.8c)$$
The additional diagrams arising from this expanded set of superheavy fermions
are:\\
\noindent D11:\qquad ${\bf 16}_1 \ {\bf - 1 - \overline{126} - 1 - 1' - 45}_X
		 \ {\bf - 16}_1,\qquad {\bf 16}_1 \ {\bf - 45}_X 
		\ {\bf - 1' - 1 - \overline{126} - 1 - 16}_1$\\
\noindent D11:\qquad ${\bf 16}_1 \ {\bf - 1 - 45}^*_Z \ {\bf - 1 - 1' - 1 - 
		\overline{126} - 1 - 16}_1$,\\[-0.1in]
\hspace*{0.9in} ${\bf 16}_1 \ {\bf - 1 - \overline{126} - 1 - 1' - 1 - 45}^*_Z
		\ {\bf - 1 - 16}_1$\\
\noindent D11:\qquad ${\bf 16}_1 \ {\bf - 1 - 45}^*_Z \ {\bf - 1 - 1' - 1 - 
		10' - 1 - 1' - 45}_X \ {\bf - 16}_1$,\\[-0.1in]
\hspace*{0.9in} ${\bf 16}_1 \ {\bf - 45}_X \ {\bf - 1' - 1 - 10' - 1 - 1' - 1 
		- 45}^*_Z \ {\bf - 1 - 16}_1$\\
\noindent D12:\qquad ${\bf 16}_1 \ {\bf - 1 - \overline{126} - 1 - 1' - 45}^*_X
		 \ {\bf - 16}_2$\\
\noindent D21:\qquad ${\bf 16}_2 \ {\bf - 45}^*_X \ {\bf - 1' - 1 - 
		\overline{126} - 1 - 16}_1$\\
\noindent D12:\qquad ${\bf 16}_1 \ {\bf - 1 - \overline{126} - 45}^*_Z \ {\bf - 
		45}_X \ {\bf - 1 - 16}_2$ \\
\noindent D21:\qquad ${\bf 16}_2 \ {\bf - 1 - 45}_X \ {\bf - 45}^*_Z \ {\bf - 
		\overline{126} - 1 - 16}_1$\\
\noindent D12:\qquad ${\bf 16}_1 \ {\bf - 1 - 45}^*_Z \ {\bf - \overline{126} - 
		45}_X \ {\bf - 1 - 16}_2$ \hfill (3.7d)\\
\noindent D21:\qquad ${\bf 16}_2 \ {\bf - 1 - 45}_X \ {\bf - \overline{126} - 
		45}^*_ Z \ {\bf - 1 - 16}_1$\\
\noindent D13:\qquad ${\bf 16}_1 \ {\bf - 1 - \overline{126} - 45}^*_Z \ {\bf - 
		1'^* - 1 - 16}_3$\\
\noindent D31:\qquad ${\bf 16}_3 \ {\bf - 1 - 1'^* - 45}^*_Z \ {\bf - 
		\overline{126} - 1 - 16}_1$\\
\noindent D13:\qquad ${\bf 16}_1 \ {\bf - 1 - 45}^*_Z \ {\bf - \overline{126} - 
		1'^* - 1 - 16}_3$\\
\noindent D31:\qquad ${\bf 16}_3 \ {\bf - 1 - 1'^* - \overline{126} - 45}^*_Z \ 
		{\bf - 1 - 16}_1$\\
\noindent D13:\qquad ${\bf 16}_1 \ {\bf - 1 - 45}^*_Z \ {\bf - 1 - 10' - 45}_X 
		\ {\bf - 16}_3$\\
\noindent D31:\qquad ${\bf 16}_3 \ {\bf - 45}_X \ {\bf - 10' - 1 - 45}^*_Z \ 
		{\bf - 1 - 16}_1$\\
\noindent M11:\qquad ${\bf 16}_1 \ {\bf - 45}_X \ {\bf - 1' - 45}^*_Z \ {\bf - 
		45}^*_Z \ {\bf - 45}_X \ {\bf - 1' - \overline{126}' - }$
		\\[-0.1in]
\hspace*{2.25in} ${\bf 1' - 45}_X \ {\bf - 45}^*_Z 
		\ {\bf - 45}^*_Z \ {\bf - 1' - 45}_X \ {\bf - 16}_1$

We thus have obtained the complete set of diagrams we shall consider for the
evaluation of the mass matrices.  Any additional diagrams for a given
$M_{Dirac}$ or $M^R$ matrix element allowed by the $U(1)_F$ family symmetry 
are of higher-order and will lead to noticeably smaller contributions to that
element than those arising from all the diagrams listed above.

\section{EVALUATION OF CONTRIBUTIONS to the MASS MATRICES}

We now turn to the evaluation of the contributions to each matrix element at
the SUSY GUT scale.  The renormalizable 3-point couplings times VEVs for 
the ${\bf 10}(5),\ {\bf 10}(\bar{5}),\ {\bf 10'}(\bar{5}'),\\{\overline{\bf
126}}(5),\ {\overline{\bf 126}}(\overline{45})$ and ${\overline{\bf 126}}'(1)$
vertices contributing to $M^U$ and $M^N$, $M^D$ and $M^E$,
$M^D$ and $M^E$, $M^U$ and $M^N$, $M^D$ and $M^E$, and $M^R$, respectively,
are labeled 
	  $$g_{10}v_u,\quad g_{10}v_d,\quad g_{10'}v'_d,\quad g_{126}w_u,
		\quad g_{126}w_d,\quad g_{126'}w' \eqno(4.1a)$$
We shall assume the superheavy fermions all get massive at the same mass
scale, so each ${\bf 1,\ 1',\ 45}_X$ or ${\bf 45}_Z$ vertex factor can be 
rescaled by the same propagator mass $M$ according to
	  $$x \equiv g_{45_X}u_{45_X}/M,\ z \equiv g_{45_Z}u_{45_Z}/M,
	    \ s \equiv g_{1}u_1/M,\ s' \equiv g_{1'}u_{1'}/M \eqno(4.1b)$$
where we have introduced a convenient short-hand notation.  In order to 
accommodate CP violation, as noted earlier after (3.5b) we introduce the 
four phases 
	  $$\phi_x,\quad \phi_z,\quad \phi_{1},\quad \phi_{1'}
	   \eqno(4.1c)$$
As a result we are led to introduce 14 independent parameters in order to 
explain the 15 quark and lepton masses and 8 quark and lepton mixing
parameters.

The contributions for each diagram then follow by moving along each fermion
line and appending the above parameters together with the coupling coefficents
spelled out in Table~I.  Alternatively, one can use the detailed computational
procedure of Mohapatra and Sakita [23] which makes explicit use of the $SU(5)$
decompositions of the $SO(10)$ matrices and fields.  We have used both 
procedures for a check in our calculations and both agree.  In the expressions 
presented below, we have evaluated the Dirac $(\psi_L)^T C^{-1}(\psi^c)_L$
and Majorana ${(\psi^c)_L}^T C^{-1}(\psi^c)_L$ matrix elements .

\noindent {\bf Leading-Order Dirac Matrix Diagrams of (3.7a):}\\[0.1in]
\noindent D33:\qquad ${\bf 16}_3 \ {\bf - 10 -} \ {\bf 16}_3$\\
\hspace*{1.2in} $M^U_{33} = M^N_{33} = g_{10}v_u,\qquad M^D_{33} = M^E_{33} = 
		g_{10}v_d$\\[0.1in]
\noindent D23:\qquad ${\bf 16}_2 \ {\bf - 10' -} \ {\bf 16}_3$\\[-0.1in]
\noindent D32:\qquad ${\bf 16}_3 \ {\bf - 10' -} \ {\bf 16}_2$\\
\hspace*{1.2in} $M^D_{23} = M^D_{32} = M^E_{23} = M^E_{32} = g'_{10}v'_d$
	\\[0.1in]
\noindent D22:\qquad ${\bf 16}_2 \ {\bf - \overline{126} -} \ {\bf 16}_2$
	\hfill (4.2a)\\
\hspace*{1.2in}	$(M^U_{22},\ M^N_{22}) = (1,\ -3)g_{126}w_u,\qquad 
	(M^D_{22},\ M^E_{22}) = (1,\ -3)g_{126}w_d$\\[0.1in]
\noindent D13:\qquad ${\bf 16}_1 \ {\bf - 45}_X \ {\bf - 10' - 45}_X \ {\bf
		- 16}_3$\\[-0.1in]
\noindent D31:\qquad ${\bf 16}_3 \ {\bf - 45}_X \ {\bf - 10' - 45}_X \ {\bf
		- 16}_1$\\
\hspace*{1.2in}	$M^D_{13} = M^D_{31} = M^E_{13} = M^E_{31} = 
	-3g_{10'}v'_d x^2 e^{2i\phi_x}$\\[0.1in]
\newpage
\noindent D12:\qquad ${\bf 16}_1 \ {\bf - 45}_Z \ {\bf - 10'^* - 45}_Z \ 
		{\bf - 16}_2$\\[-0.1in]
\noindent D21:\qquad ${\bf 16}_2 \ {\bf - 45}_Z \ {\bf - 10'^* - 45}_Z \ 
		{\bf - 16}_1$\\
\hspace*{1.2in} $M^D_{12} = M^D_{21} = -4g_{10'}v'_d z^2
		e^{2i\phi_z},\qquad M^E_{12} = M^E_{21} = 0$\\[0.3in]
\noindent {\bf Leading-Order Majorana Matrix Diagrams of (3.7b):}\\[0.1in]
\noindent M33:\qquad ${\bf 16}_3 \ {\bf - 1 - \overline{126}' - 1 - 16}_3$\\
\hspace*{1.2in} $M^R_{33} = g_{126'}w' s^2 e^{2i\phi_1}$\\[0.1in]
\noindent M23:\qquad ${\bf 16}_2 \ {\bf - 1 - 45}_X\ {\bf - 1' - 
		\overline{126}' - 1 - 16}_3$\\[-0.1in]
\noindent M32:\qquad ${\bf 16}_3 \ {\bf - 1 - \overline{126}' - 1' - 45}_X
		\ {\bf - 1 - 16}_2$\\
\hspace*{1.2in} $M^R_{23} = M^R_{32} = 5g_{126'}w' xs^2s'
		e^{i(\phi_x + \phi_1 + \phi_{1'})}$\\[0.1in]
\noindent M22:\qquad ${\bf 16}_2 \ {\bf - 1 - 45}_X\ {\bf - 1' - 
		\overline{126}' - 1' - 45}_X\ {\bf - 1 - 16}_2$\\
\hspace*{1.2in} $M^R_{22} = 25g_{126'}w'(xss')^2
		e^{2i(\phi_x + \phi_1 + \phi_{1'})}$\\[0.1in]
\noindent M13:\qquad ${\bf 16}_1 \ {\bf - 45}_X\ {\bf - 1' - 1- 45}_Z
		\ {\bf - 1' - \overline{126}' - 1 - 16}_3$\hfill 
		(4.2b)\\[-0.1in]
\noindent M31:\qquad ${\bf 16}_3 \ {\bf - 1 - \overline{126}' - 1' - 45}_Z
		\ {\bf - 1 - 1' - 45}_X\ {\bf - 16}_1$\\
\hspace*{1.2in} $M^R_{13} = M^R_{31} = 30g_{126'}w' xzs^2 s'^2
		e^{i(\phi_x + \phi_z + 2\phi_1
		+ 2\phi_{1'})}$\\[0.1in]
\noindent M12:\qquad ${\bf 16}_1 \ {\bf - 45}_X\ {\bf - 1' - 1- 45}_Z
		\ {\bf - 1' - \overline{126}' - 1' - 45}_X\ {\bf - 1 - 16}_2$
		\\[-0.1in]
\noindent M21:\qquad ${\bf 16}_2 \ {\bf - 1 - 45}_X\ {\bf - 1' - 
		\overline{126}' - 1' - 45}_Z\ {\bf - 1 - 1' - 45}_X
		\ {\bf - 16}_1$\\
\hspace*{1.2in} $M^R_{12} = M^R_{21} = 150g_{126'}w' x^2 z s^2 s'^3 
		e^{i(2\phi_x + \phi_z + 2\phi_1
		+ 3 \phi_{1'})}$\\[0.1in]
\noindent M11:\qquad ${\bf 16}_1 \ {\bf - 45}_X\ {\bf - 1' - 1- 45}_Z
		\ {\bf - 1' - \overline{126}' - 1' - 45}_Z
		\ {\bf - 1 - 1' - 45}_X {\bf - 16}_1$\\
\hspace*{1.2in} $M^R_{11} = 900g_{126'}w'(xzss'^2)^2
		e^{2i(\phi_x + \phi_z + \phi_1
		+ 2\phi_{1'})}$\\[0.3in]
\noindent {\bf Higher-Order Diagrams listed in (3.7c) from Minimal Set:}
	\\[0.1in]
\noindent D11:\qquad ${\bf 16}_1 \ {\bf - 45}_X\ {\bf - 1' - 1 - 10' - 1 
		- 1' - 45}_X\ {\bf - 16}_1$\\
\hspace*{1.2in} $M^D_{11} = M^E_{11} = -3g_{10'}v'_d(xss')^2
		e^{2i(\phi_x + \phi_1 + \phi_{1'})}$\\[0.1in]
\newpage
\noindent D22:\qquad ${\bf 16}_2 \ {\bf - 45}_Z\ {\bf - 10'^* - 1'^* - 16}_2,
		\qquad {\bf 16}_2 \ {\bf - 1'^* - 10'^* - 45}_Z\ {\bf - 16}_2$\\
\hspace*{1.2in} $M^D_{22} = M^E_{22} = -3g_{10'}v'_d zs'
		e^{i(\phi_z - \phi_{1'})}$\hfill (4.2c)\\[0.05in]
\noindent M23:\qquad ${\bf 16}_2 \ {\bf - 45}_X^*\ {\bf - 1' - 1 - 45}_Z
		\ {\bf - 1' - \overline{126}' - 1 - 16}_3$\\[-0.1in]
          M32:\qquad ${\bf 16}_3 \ {\bf - 1 - \overline{126}' - 1' - 45}_Z
		\ {\bf - 1 - 1' - 45}_X^*\ {\bf - 16}_2$\\
\hspace*{1.2in} $M^R_{23} = M^R_{32} = 30g_{126'}w' xzs^2 s'^2
		e^{i(-\phi_x + \phi_z + 2\phi_1 + 
		2\phi_{1'})}$\\[0.2in]
\noindent {\bf Higher-Order Diagrams of (3.7d) from the Expanded Set:}
	\\[0.1in]
\noindent D11:\qquad ${\bf 16}_1 \ {\bf - 1 - \overline{126} - 1 - 1' - 45}_X
		 \ {\bf - 16}_1,\qquad {\bf 16}_1 \ {\bf - 45}_X 
		\ {\bf - 1' - 1 - \overline{126} - 1 - 16}_1$\\
\hspace*{1.2in} $(M^U_{11},\ M^N_{11}) = (2,\ -6)g_{126}w_u xs^2 s'
		e^{i(\phi_x + 2\phi_1 + \phi_{1'})}$\\
\hspace*{1.2in} $(M^D_{11},\ M^E_{11}) = (-2,\ 6)g_{126}w_d xs^2s'
		e^{i(\phi_x + 2\phi_1 + \phi_{1'})}$\\[0.05in]
\noindent D11:\qquad ${\bf 16}_1 \ {\bf - 1 - 45}^*_Z \ {\bf - 1 - 1' - 1 - 
		\overline{126} - 1 - 16}_1$,\\[-0.1in]
\hspace*{0.9in} ${\bf 16}_1 \ {\bf - 1 - \overline{126} - 1 - 1' - 1 - 45}^*_Z
		\ {\bf - 1 - 16}_1$\\
\hspace*{1.2in} $(M^U_{11},\ M^N_{11}) = (3,\ -9)g_{126}w_u zs^4 s'
		e^{i(-\phi_z + 4\phi_1 + \phi_{1'})}$\\
\hspace*{1.2in} $(M^D_{11},\ M^E_{11}) = (-3,\ 9)g_{126}w_d zs^4 s'
		e^{i(-\phi_z + 4\phi_1 + \phi_{1'})}$\\[0.05in]
\noindent D11:\qquad ${\bf 16}_1 \ {\bf - 1 - 45}^*_Z \ {\bf - 1 - 1' - 1 - 
		10' - 1 - 1' - 45}_X \ {\bf - 16}_1$,\\[-0.1in]
\hspace*{0.9in} ${\bf 16}_1 \ {\bf - 45}_X \ {\bf - 1' - 1 - 10' - 1 - 1' - 1 
		- 45}^*_Z \ {\bf - 1 - 16}_1$\\
\hspace*{1.2in} $(M^D_{11},\ M^E_{11}) = (-7,\ -3)g_{10'}v'_d xzs^4 s'^2
		e^{i(\phi_x - \phi_z + 4\phi_1 +
		2\phi_{1'})}$\\[0.05in]
\noindent D12:\qquad ${\bf 16}_1 \ {\bf - 1 - \overline{126} - 1 - 1' - 45}^*_X
		 \ {\bf - 16}_2$\\
\hspace*{1.2in} $(M^U_{12},\ M^N_{12}) = (1,\ -15)g_{126}w_u xs^2s'
		e^{i(-\phi_x + 2\phi_1 + \phi_{1'})}$\\
\hspace*{1.2in} $M^D_{12} = M^E_{12} = -3g_{126}w_d xs^2s'
		e^{i(-\phi_x + 2\phi_1 + \phi_{1'})}$\\[0.05in]
\noindent D21:\qquad ${\bf 16}_2 \ {\bf - 45}^*_X \ {\bf - 1' - 1 - 
		\overline{126} - 1 - 16}_1$\\
\hspace*{1.2in} $(M^U_{21},\ M^N_{21}) = (1,\ 9)g_{126}w_u xs^2s'
		e^{i(-\phi_x + 2\phi_1 + \phi_{1'})}$\\
\hspace*{1.2in} $(M^D_{21},\ M^E_{21}) = (1,\ 9)g_{126}w_d xs^2s'
		e^{i(-\phi_x + 2\phi_1 + \phi_{1'})}$\\[0.05in]
\noindent D12:\qquad ${\bf 16}_1 \ {\bf - 1 - \overline{126} - 45}^*_Z \ {\bf - 
		45}_X \ {\bf - 1 - 16}_2$\\
\hspace*{1.2in} $(M^U_{12},\ M^N_{12}) = (2,\ -90)g_{126}w_u xzs^2
		e^{i(\phi_x - \phi_z + 2\phi_1)}$\\
\hspace*{1.2in} $M^D_{12} = 12g_{126}w_d xzs^2
		e^{i(\phi_x - \phi_z + 2\phi_1)},
		\qquad M^E_{12} = 0$\\
\newpage
\noindent D21:\qquad ${\bf 16}_2 \ {\bf - 1 - 45}_X \ {\bf - 45}^*_Z \ {\bf - 
		\overline{126} - 1 - 16}_1$\\
\hspace*{1.2in} $(M^U_{21},\ M^N_{21}) = (1,\ -27)g_{126}w_u xzs^2
		e^{i(\phi_x - \phi_z + 2\phi_1)}$\\
\hspace*{1.2in} $(M^D_{21},\ M^E_{21}) = (1,\ -27)g_{126}w_d xzs^2
		e^{i(\phi_x - \phi_z + 2\phi_1)}$\\[0.1in]
\noindent D12:\qquad ${\bf 16}_1 \ {\bf - 1 - 45}^*_Z \ {\bf - \overline{126} - 
		45}_X \ {\bf - 1 - 16}_2$\hfill (4.2d)\\
\hspace*{1.2in} $(M^U_{12},\ M^N_{12}) = (1,\ 45)g_{126}w_u xzs^2
		e^{i(\phi_x - \phi_z + 2\phi_1)}$\\
\hspace*{1.2in} $(M^D_{12},\ M^E_{12}) = (-3,\ 9)g_{126}w_d xzs^2
		e^{i(\phi_x - \phi_z + 2\phi_1)}$\\[0.1in]
\noindent D21:\qquad ${\bf 16}_2 \ {\bf - 1 - 45}_X \ {\bf - \overline{126} - 
		45}^*_ Z \ {\bf - 1 - 16}_1$\\
\hspace*{1.2in} $(M^U_{21},\ M^N_{21}) = (2,\ 54)g_{126}w_u xzs^2
		e^{i(\phi_x - \phi_z + 2\phi_1)}$\\
\hspace*{1.2in} $M^D_{21} = -4g_{126}w_d xzs^2
		e^{i(\phi_x - \phi_z + 2\phi_1)},
		\qquad M^E_{21} = 0$\\[0.1in]
\noindent D13:\qquad ${\bf 16}_1 \ {\bf - 1 - \overline{126} - 45}^*_Z \ {\bf - 
		1'^* - 1 - 16}_3$\\
\hspace*{1.2in} $(M^U_{13},\ M^N_{13}) = (2,\ -18)g_{126}w_u zs^2 s'
		e^{i(-\phi_z + 2\phi_1 - \phi_{1'})}$\\
\hspace*{1.2in} $M^D_{13} = -4g_{126}w_d zs^2 s'
		e^{i(-\phi_z + 2\phi_1 - \phi_{1'})},
		\qquad M^E_{13} = 0$\\[0.1in]
\noindent D31:\qquad ${\bf 16}_3 \ {\bf - 1 - 1'^* - 45}^*_Z \ {\bf - 
		\overline{126} - 1 - 16}_1$\\
\hspace*{1.2in} $(M^U_{31},\ M^N_{31}) = (1,\ 9)g_{126}w_u zs^2 s'
		e^{i(-\phi_z + 2\phi_1 - \phi_{1'})}$\\
\hspace*{1.2in} $(M^D_{31},\ M^E_{31}) = (1,\ 9)g_{126}w_d zs^2 s'
		e^{i(-\phi_z + 2\phi_1 - \phi_{1'})}$\\[0.1in]
\noindent D13:\qquad ${\bf 16}_1 \ {\bf - 1 - 45}^*_Z \ {\bf - \overline{126} - 
		1'^* - 1 - 16}_3$\\
\hspace*{1.2in} $(M^U_{13},\ M^N_{13}) = (1,\ 9)g_{126}w_u zs^2 s'
		e^{i(-\phi_z + 2\phi_1 - \phi_{1'})}$\\
\hspace*{1.2in} $(M^D_{13},\ M^E_{13}) = (1,\ 9)g_{126}w_d zs^2 s'
		e^{i(-\phi_z + 2\phi_1 - \phi_{1'})}$\\[0.1in]
\noindent D31:\qquad ${\bf 16}_3 \ {\bf - 1 - 1'^* - \overline{126} - 45}^*_Z \ 
		{\bf - 1 - 16}_1$\\
\hspace*{1.2in} $(M^U_{31},\ M^N_{31}) = (2,\ -18)g_{126}w_u zs^2 s'
		e^{i(-\phi_z + 2\phi_1 - \phi_{1'})}$\\
\hspace*{1.2in} $M^D_{31} = -4g_{126}w_d zs^2 s'
		e^{i(-\phi_z + 2\phi_1 - \phi_{1'})},
		\qquad M^E_{31} = 0$\\[0.1in]
\noindent D13:\qquad ${\bf 16}_1 \ {\bf - 1 - 45}^*_Z \ {\bf - 1 - 10' - 45}_X 
		\ {\bf - 16}_3$\\
\hspace*{1.2in} $M^D_{13} = M^E_{13} = -3g_{10'}v'_d xzs^2
		e^{i(\phi_x - \phi_z + 2\phi_1)}$\\[0.1in]
\noindent D31:\qquad ${\bf 16}_3 \ {\bf - 45}_X \ {\bf - 10' - 1 - 45}^*_Z \ 
		{\bf - 1 - 16}_1$\\
\hspace*{1.2in} $M^D_{31} = -4g_{10'}v'_d xzs^2
		e^{i(\phi_x - \phi_z + 2\phi_1)},
		\qquad M^E_{31} = 0$\\[0.1in]
\noindent M11:\qquad ${\bf 16}_1 \ {\bf - 45}_X \ {\bf - 1' - 45}^*_Z \ {\bf - 
		45}^*_Z \ {\bf - 45}_X \ {\bf - 1' - \overline{126}' - }$
		\\[-0.1in]
\hspace*{2.25in} ${\bf 1' - 45}_X \ {\bf - 45}^*_Z 
		\ {\bf - 45}^*_Z \ {\bf - 1' - 45}_X \ {\bf - 16}_1$\\
\hspace*{1.2in} $M^R_{11} = (900)^2 g_{126'}w'(xzs')^4
		e^{4i(\phi_x - \phi_z + \phi_{1'})}$\\[0.1in]

An interesting observation which can be drawn from the Majorana contributions
in (4.2b) is that the matrix in leading order has a geometrical texture as 
given in (2.4c) with 
	$$M^R_{22} \simeq {5\over{6}}\sqrt{M^R_{11}M^R_{33}} \eqno(4.3)$$
provided $x \simeq z$.  
In fact, this observation served as an important guide in 
our construction of the Majorana neutrino matrix and suggested the relative
roles played by the ${\bf 45}_X$ and ${\bf 45}_Z$ Higgs fields.

\section{QUANTITATIVE RESULTS for the $SO(10) \times U(1)_F$ MODEL}

Finally we attempt to select a set of values for the 14 input parameters of 
(4.1a,b,c) which will accurately reproduce the input data in (2.1a,b) and 
(2.3a,b) used for our bottom-up approach.  As noted earlier, the minimal
set of superheavy fermions and their mirror partners found in (3.8b) yield
unsatisfactory results: $m_u = m_{\nu_e} = 0,\ m_e = 0.006$ MeV and 
$m_{\nu_{\mu}} = m_{\nu_{\tau}} = 0.089$ eV.  The problem can be traced to 
the zero or tiny values of D11.  By expanding the set of
superheavy fermions to include those in (3.8c), on the other hand, excellent
results can be found as shown below.  

One particularly good numerical choice for the parameters at the SUSY GUT scale
is given by\\
	$$\begin{array}{rlrlrlrl}
	 g_{10}v_u &= 120.3, &g_{10}v_d &= 2.46, &g_{10'}v'_d 
		&\multicolumn{2}{l}{= 0.078\ {\rm GeV}}\nonumber\\
	 g_{126}w_u &= 0.314, &g_{126}w_d &= - 0.037, &g_{126'}w' 
		&\multicolumn{2}{l}{= 0.8 \times 10^{16}\ {\rm GeV}}\cr
	 g_{45_X}u_{45_X}/M &= 0.130, &g_{45_Z}u_{45_Z}/M &= 0.165, 
		&g_1u_1/M &= 0.56, &g_{1'}u'_1/M &= - 0.026\cr
	 \phi_x &= 35^o, &\phi_z &\multicolumn{2}{l}{= \phi_{1} = 
		\phi_{1'} = - 5^o} \cr
	  \end{array} \eqno(5.1)$$
which reduces the number of independent parameters from 14 to 12.  In fact,
the only large phase angle is that for $\phi_x$.  As seen from (4.2a),
this is in agreement with our earlier conclusion from the bottom-up 
phenomenological results [8] that essentially only the Dirac D13 and D31 matrix 
elements are complex.   The mass matrices at the SUSY GUT scale are then 
numerically equal to 
$$\eqalignno{\noindent
	M^U&= \left(\matrix{ -0.0010 - 0.0001i& 0.0053 + 0.0034i& -0.0013\cr
	0.0053 + 0.0034i& 0.314 & 0 \cr
	-0.0013 & 0 & 120.3 \cr}\right)&(5.2a)\cr
\noindent M^D&= \left(\matrix{ -0.0001 & -0.0104 + 0.0004i& 
	-0.0029 - 0.0045i \cr
	-0.0077 + 0.0018i & -0.036 & 0.078 \cr
	-0.0033 - 0.0048i & 0.078 & 2.460 \cr}\right)&(5.2b)\cr
\noindent M^N&= \left(\matrix{ 0.0030 + 0.0003i & -0.079 - 0.051i & 0.0038 \cr
	0.048 + 0.031i & -0.942 & 0 \cr
	0.0038 & 0 & 120.3 \cr}\right)&(5.2c)\cr
\noindent M^E&= \left(\matrix{ 0.0004  & -0.0020 - 0.0010i &
	-0.0023 - 0.0045i \cr
	0.0060 + 0.0031i & 0.112 & 0.078 \cr
	-0.0009 - 0.0037i & 0.078 & 2.460 \cr}\right)&(5.2d)\cr
\noindent M^R&= \left(\matrix{(-.069 + .640i)\times 10^{9}&
	(-.141 - .119i)\times 10^{11}& (.108 + .019i)\times 10^{13}\cr
	(-.141 - .119i)\times 10^{11}& (.461 + .549i)\times 10^{12}&
	(-.393 - .155i)\times 10^{14}\cr
	(.108 + .019i)\times 10^{13}& (-.393 - .155i)\times 10^{14}&
	(.247 - .044i)\times 10^{16}\cr}\right)&(5.2e)\cr}$$
in units of GeV.  By using the seesaw formula [16], we find for 
the light neutrino matrix at the SUSY GUT scale
$$\begin{array}{rl}
\noindent M^{N_{eff}}&\simeq -M^N(M^R)^{-1}{M^{N^T}}\nonumber\\[0.1in]
       &= \left(\matrix{(.027 - .238i)\times 10^{-3}& 
	(-.109 - .199i)\times 10^{-2}& (-.037 + .512i)\times 10^{-2}\cr
	(-.109 - .199i)\times 10^{-2}& (-.232 - .088i)\times 10^{-1}&
	(.258 + .435i)\times 10^{-1}\cr
	(-.037 + .512i)\times 10^{-2}& (.258 + .435i)\times 10^{-1}&
	-.001 - .112i\cr}\right)
	\end{array} \eqno(5.2f)$$
in electron-Volts. Again we emphasize the Dirac mass matrix elements 
appear in the form 
$\psi_{iL}^T C^{-1}M(\psi^c)_{jL}$, while the Majorana matrix elements refer to 
$(\psi^c)_{iL}^T C^{-1}M(\psi^c)_{jL}$ with $\psi_{iL}$ and $(\psi^c)_{jL}$ 
each a member 
of one of the three families of ${\bf 16}$'s.  Identical contributions also
arise from the transposed Dirac matrices and the right-handed Majorana 
matrix.  As such, the true Yukawa couplings $G_Y$ are just half the values of 
the $g_Y$'s appearing in (4.1a,b).

The masses at the GUT scale can then be found by calculating the eigenvalues 
of the Hermitian product $MM^{\dagger}$ in each case, while the mixing matrices
$V_{CKM}$ and $V_{lept}$ can be calculated with the projection operator
technique of Jarlskog [24].  After evolving these quantities to the low scale,
we find in the quark sector
$$\begin{array}{rlrl}
        m_u(1 {\rm GeV})&= 5.0\ (5.1)\ {\rm MeV},& \qquad m_d(1 {\rm GeV})&= 
		7.9\ (8.9)\ {\rm MeV}\nonumber\\
        m_c(m_c)&= 1.27\ (1.27)\ {\rm GeV},& \qquad m_s(1 {\rm GeV})&= 169
		\ (175)\ {\rm MeV}\cr
        m_t(m_t)&= 150\ (165)\ {\rm GeV},& \qquad m_b(m_b)&= 4.09\ (4.25)
		\ {\rm GeV}\cr
  \end{array}\eqno(5.3a)$$
where we have indicated the preferred values in parentheses.  The 
mixing matrix is given by 
$$V_{CKM} = \left(\matrix{0.972 & 0.235 & 0.0037e^{-i124^o}\cr
                -0.235 & 0.971 & 0.041\cr
                0.012 & -0.039 & 0.999\\[-0.15in]\cr
		\quad -0.003i & \quad -0.001i & \cr}\right) \eqno(5.3b)$$
Note that $V_{cb} = 0.041$ and $|V_{ub}/V_{cb}| = 0.090$ with the CP-violating 
phase $\delta = 124^o$, while $m_d/m_u = 1.59$ and
$m_s/m_d = 21.3$, cf. [12, 13].  These results should be compared with our 
central starting input values given in (2.1a,b).  

In the lepton sector we obtain
$$\begin{array}{rlrl}
        m_{\nu_e}&= 0.10\ (?) \times 10^{-4}\ {\rm eV},& \qquad m_e&= 
		0.43\ (0.511)\ {\rm MeV}\nonumber\\
        m_{\nu_{\mu}}&= 0.29\ (0.25) \times 10^{-2}\ {\rm eV},& \qquad m_{\mu}
		&= 103\ (105.5)\ {\rm MeV}\cr
        m_{\nu_{\tau}}&= 0.12\ (0.10)\ {\rm eV},& \qquad m_{\tau}&= 1.777
		\ (1.777)\ {\rm GeV}\cr \end{array}\eqno(5.4a)$$
and 
$$V_{lept} = \left(\matrix{0.998 & 0.049 & 0.039e^{-i121^o}\cr
                -0.036  & 0.875 & 0.483\qquad\cr
                0.042   & -0.482 & 0.875\\[-0.15in]\cr
	        \quad -0.037i & \quad -0.002i & \cr}\right) \eqno(5.4b)$$
which should be compared with the input values in (2.3a,b).  
The heavy Majorana neutrino masses are 
	$$M^R_1 = 0.63 \times 10^9\ {\rm GeV},\qquad M^R_2 = 0.37 \times 
		10^{11}\ {\rm GeV},\qquad M^R_3 = 0.25 \times 10^{16}\ {\rm
		GeV} \eqno(5.4c)$$
The neutrino masses and mixings are in the correct ranges to explain the 
nonadiabatic solar neutrino depletion with small mixing [5] and the atmospheric
neutrino depletion with large mixing [6]:
	$$\begin{array}{rlrl}
	  \delta m^2_{12}&= 8.5 \times 10^{-6}\ {\rm eV^2},&\qquad 
		\sin^2 2\theta_{12}&= 0.0062\nonumber\\
	  \delta m^2_{23}&= 1.4 \times 10^{-2}\ {\rm eV^2},&\qquad 
		\sin^2 2\theta_{23}&= 0.71 \cr \end{array} \eqno(5.5)$$

For our analysis, the SUSY GUT scale at which the gauge and Yukawa couplings 
unify was chosen to be $\Lambda = 1.2 \times 10^{16}$ GeV.  From (3.4a) and 
(5.2a,b,c,d) 
we find that $g_{10} = 0.69$.  It is interesting to note that if we equate the 
$SO(10)$-breaking and lepton number-breaking VEV, $w'$, with $\Lambda$, we 
find $g_{126'} = 0.67 \simeq g_{10}$.  Taking into account the remark 
following (5.2e), we note the true Yukawa couplings are $G_{10} \simeq 
G_{126'} \simeq 0.33$.  If we further equate $g_1 = g_{10} \simeq g_{126'}$, 
and $u_1 = \Lambda$ for the $U(1)_F$-breaking VEV, we find $M = 1.5 \times
10^{16}$ GeV for the masses of the superheavy fermions which condense 
with their mirrors.  These values are all very reasonable.

The ${\bf 45}_X$ and ${\bf 45}_Z$ VEVs appear at nearly
the same scale, $2.8 \times 10^{15}$ and $3.5 \times 10^{15}$ GeV respectively,
if one assumes the same Yukawa coupling as above.  On the other hand, if these
VEVs appear at the unification scale $\Lambda$ the corresponding Yukawa 
couplings
are smaller than those found above.  In either case, a consequence of their
non-orthogonal breakings is that $SU(5)$ is broken down to $SU(3)_c \times
SU(2)_L \times U(1)_Y$ at the scale in question.  No further breaking is 
required until the electroweak scale and the SUSY-breaking scale are reached.  

\section{SUMMARY}

Our starting point for this research has been based on the results obtained
from a bottom-up approach proposed previously by us to obtain mass matrices
at the SUSY GUT scale based on a complete set of data inputted at the 
low scales.  In particular we have used the known quark and charged lepton 
masses and CKM mixing matrix together with the neutrino masses and mixings 
based on particular neutrino scenarios.  The masses and mixing
matrices were evolved to the SUSY GUT scale where the mass matrices can be 
constructed by use of Sylvester's theorem.   By varying the bases and the 
signs of the mass eigenvalues, we looked for simple textures for the mass 
matrices such that each matrix element involved as few $SO(10)$ Higgs 
representations as possible. The neutrino scenario examined which appeared to 
yield the simplest structure involved the MSW nonadiabatic depletion of the 
solar electron-neutrinos together with the observed depletion of atmospheric 
muon-neutrinos by oscillations into tau-neutrinos.

In this paper we have constructed an $SO(10) \times U(1)_F$ model of the 
Yukawa interactions which neatly reproduces the desired $SO(10)$ textures for 
the quark and lepton mass matrices for this preferred neutrino scenario.
The observed features include the following: 

\indent (i) The Abelian $U(1)_F$ family symmetry group singles out a rather 
simple set of tree diagrams which determines the texture
of the generic Dirac and Majorana mass matrices, while the $SO(10)$ group
relates corresponding matrix elements of the up, down, neutrino and charged
lepton Dirac matrices to each other. \\
\indent (ii) The dominant second and third family Yukawa interactions are 
renormalizable and arise through couplings with Higgs in the ${\bf 10,\ 10'}$ 
and $\bf \overline{126}$ representations of $SO(10)$.  The remaining Yukawa
interactions are of higher order and require couplings of Higgs in the
${\bf \overline{126}}',\ {\bf 1,\ 1',\ 45}_X$ and ${\bf 45}_Z$ representations 
which acquire VEVs near the SUSY GUT scale.\\
\indent (iii) The Higgs 
which acquire high scale VEVs break the $SO(10) \times U(1)_F$ symmetry down 
to the $SU(3)_c \times SU(2)_L \times U(1)_Y$ standard model symmetry in two 
stages through the $SU(5)$ subgroup.\\
\indent (iv) Although this non-minimal supersymmetric model involves several 
Higgs representations,
the runnings of the Yukawa couplings from the GUT scale to the low-energy 
SUSY-breaking scale are controlled mainly by the contributions from the 
${\bf 10}$, as in the minimal supersymmetric standard model.\\
\indent (v) The complete set of low scale VEVs which contribute to the fermion
masses are ${\bf 10}(5),\ {\bf 10}(\bar{5}),\ {\bf 10}'(\bar{5}'),\ \overline{
\bf 126}(5)$ and $\overline{\bf 126}(\overline{45})$ in the $SO(10)(SU(5))$
notation.  These Higgs correspond to the minimum number required in $SO(10)$
models which lead to the successful Georgi - Jarlskog relations [3].  Most
of these models, however, do not include neutrino mass matrices.\\
\indent (vi) In terms of 12 input parameters, 15 masses (including the 
heavy Majorana masses) and 8 mixing parameters emerge.  The Yukawa couplings 
and the Higgs VEVs are numerically feasible and successfully
correlate all the quark and lepton masses and mixings in the scenario which
incorporates the nonadiabatic solar neutrino and atmospheric neutrino 
depletion effects.\\
\indent (vii) The right-handed Majorana neutrino matrix has a nearly geometrical
texture leading to heavy Majorana neutrino masses spread over seven orders of
magnitude as given in (5.4c).  In fact, it is the highly geometrical structure
of the Majorana matrix which accounts for the nearly maximal mixing of the 
$\nu_{\mu}$ and $\nu_{\tau}$, rather than sizable mixing in the Dirac 
sector [25].

With the model as presented, the $U(1)_F$ current is anomalous, since the 
$[U(1)_F]^3$ triangle anomaly does not vanish.  It is possible to cancel this
anomaly, however, by the addition of two $SO(10)$ singlet neutral fermions, 
$n_L$ and $(n^c)_L$, both with $U(1)_F$ charges of -12.  By introducing 
another Higgs singlet representation which develops a GUT scale VEV, one can 
arrange that one of the new neutrinos remains massless while the other becomes
superheavy.  Alternatively, it is possible to cancel such an anomaly through
the Green-Schwarz mechanism [22] provided the model can be derived from string
theory.

Studies are underway to examine what effects small mixings of such
a light sterile neutrino with the three families of light neutrinos will have 
on the neutrino spectrum and will be reported elsewhere.  Work is also underway
to construct a superpotential for the model presented here.\\[0.5in]

The authors thank the Fermilab Theoretical Physics Department for the 
opportunity to participate in the 1994 and 1995 Summer Visitor Programs during
which time the research reported here was initiated and completed. 
Useful discussions with Joseph Lykken and Rabindra Mohapatra about various 
aspects of this work are gratefully acknowledged.  This research was supported 
in part by the U. S. Department of Energy.
\newpage 
\begin{reflist}
\item	U. Amaldi, W. de Boer, and H. Furstenau, Phys. Lett. B {\bf 260}, 447
	(1991); J. Ellis, S. Kelley, and D. V. Nanopoulos, ibid. {\bf 260},
	131 (1991); P. Langacker and M. Luo, Phys. Rev. D {\bf 44}, 817 (1991).

\item	S. Weinberg, Proceedings of Festschrift in honor of I. I. Rabi,
	New York Academy of Science;  F. Wilczek and A. Zee, Phys. Lett.
	B {\bf 70}, 418 (1977), erratum, ibid. B {\bf 72}, 504 (1978);
	H. Fritzsch, Phys. Lett. B {\bf 70}, 436 (1977).

\item   H. Georgi and C. Jarlskog, Phys. Lett. B {\bf 86}, 297 (1979);
 	J. A. Harvey, P. Ramond, and D. B. Reiss, Phys. Lett. B {\bf 92}, 309
	(1980); H. Arason, D. Castan\~{n}o, B. Keszthelyi, S. Mikaelian,
	E. Piard, P. Ramond, and B. Wright, Phys. Rev. Lett. {\bf 67}, 2933 
	(1991); Phys. Rev. D {\bf 46}, 3945 (1992); 
	S. Dimopoulos, L. J. Hall, and S. Raby, Phys. Rev. Lett.
	{\bf 68}, 1984 (1992);  Phys. Rev. D {\bf 45}, 4192 (1992); 
	{\bf 46}, R4793 (1992); {\bf 47}, R3702 (1993);  G. F. Giudice,
	Mod. Phys. Lett. A {\bf 7}, 2429 (1992); H. Arason, D. Castan\~{n}o,
	P. Ramond and E. Piard, Phys. Rev. D {\bf 47}, 232 (1993);
	P. Ramond, R. G. Roberts, and G. G. Ross, Nucl. Phys. {\bf B406},
	19 (1993); A. Kusenko and R. Shrock, Phys. Rev. D {\bf 49}, 4962 (1994);
	K. S. Babu and R. N. Mohapatra, Phys. Rev. Lett. {\bf 74}, 2418 (1995).

\item   S. P. Mikheyev and A. Yu Smirnov, Yad Fiz. {\bf 42}, 1441 (1985)
        [Sov. J. Nucl. Phys. {\bf 42}, 913 (1986)]; Zh. Eksp. Teor.
        Fiz. {\bf 91}, 7 (1986) [Sov. Phys. JETP {\bf 64}, 4 
        (1986)]; Nuovo Cimento {\bf 9C}, 17 (1986); L. Wolfenstein,
        Phys. Rev. D {\bf 17}, 2369 (1978); {\bf 20}, 2634 (1979).
 
\item	R. Davis et al., Phys. Rev. Lett. {\bf 20}, 1205 (1968); in {\it 
	Neutrino '88}, ed. J. Schnepp et al. (World Scientific, 1988);
	K. Hirata et al., Phys. Rev. Lett. {\bf 65}, 1297, 1301 (1990);
        P. Anselmann et al., Phys. Lett. B {\bf 327}, 377, 390 (1994); 
	Dzh. N. Abdurashitov et al., Phys. Lett. B {\bf 328}, 234 (1994).

\item   K. S. Hirata et al., Phys. Lett. B {\bf 280}, 146 (1992); and {\bf
	283}, 446 (1992); R. Becker-Szendy et al., Phys. Rev. Lett. {\bf 69},
	(1992) and Phys. Rev. D {\bf 46}, 3720 (1992);
	W. W. M. Allison et al., Report No. ANL-HEP-CP-93-32; Y. Fukuda
	et al., Phys. Lett. B {\bf 335}, 237 (1994).
 
\item   Q. Shafi and F. Stecker, Phys. Rev. Lett. {\bf 53}, 1292 (1984);
        Ap. J. {\bf 347}, 575 (1989).
 
\item	C. H. Albright and S. Nandi, Phys. Rev. Lett. {\bf 73}, 930 (1994);
	Phys. Rev. D {\bf 52}, 410 (1995); Report No. Fermilab-PUB-94/119-T 
	and OSU Report No. 289 (to be published); C. H. Albright, in 
	Proceedings of the Yukawa Conference held at the University of 
	Florida, February 1994 (International Press, Cambridge, Massachusetts,
	1995).

\item	F. R. Grantmakher, {\it Theory of Matrices}, (Chelsa Publishing
	Company, New York, 1959).

\item	C. H. Albright and S. Nandi, Report no. FERMILAB-PUB-95/107-T and 
	OSU Report No. 303, to be published.

\item   S. Dimopoulos, Phys. Lett. B {\bf 129}, 417 (1983).

\item	S. Naculich, Phys. Rev. D {\bf 48}, 5293 (1993).

\item   Particle Data Group, M. Aguilar-Benitez et al., Phys. Rev. D {\bf 50}, 
	1173 (1994).

\item   J. Gasser and H. Leutwyler, Phys. Rep. C {\bf 87}, 77 (1982).

\item   A. Kusenko, Phys. Lett. B {\bf 284}, 390 (1992).
 
\item   M. Gell-Mann, P. Ramond, and R. Slansky, in {\it Supersymmetry}, 
        edited by P. Van Nieuwenhuizen and D. Z. Freedman 
        (North-Holland, Amsterdam, 1979); T. Yanagida, Prog. Theor.
        Phys. {\bf B 315}, 66 (1978).
 
\item   E. H. Lemke, Mod. Phys. Lett. A {\bf 7}, 1175 (1992).

\item   The form of $M^R$ presented here differs somewhat from that in Ref.
	[1], for more recent data on the atmospheric depletion effect was
	taken into account.
 
\item	For recent use of $U(1)_F$ symmetry to generate patterns of fermion
	mass matrices, see L. Ibanez and G. G. Ross, Phys. Lett. B {\bf 332},
	100 (1994); P. Binetruy and P. Ramond, Phys. Lett. B {\bf 350}, 49 
	(1995); H. Dreiner, G. K. Leontaris , S. Lola, and G. G. Ross, Nucl. 
	Phys. B {\bf 436}, 461 (1995); V. Jain and R. Shrock, Report No. 
	ITP-SB-94-55.

\item	For reviews, cf. P. Fayet and S. Ferrara, Phys. Rep. C {\bf 32}, 249
	(1977); P. Van Niewenhuizen, ibid. {\bf 68}, 189 (1981);  J. Wess and
	J. Bagger, {\it Supersymmetry and Supergravity}, Princeton University
	Press, Princeton, New Jersey (1983); P. Nath, R. Arnowitt and A. H.
	Chamseddine, Applied N = 1 Supergravity, World Scientific, Singapore
	(1983); H. P. Nilles, Phys. Rep. C {\bf 110}, 1 (1984).

\item	Jorge L. Lopez and D. V. Nanopoulos, Proceedings of 15th Johns 
	Hopkins Workshop, (1991), p. 277.

\item   M. B. Green and J. H. Schwarz, Phys. Lett. B {\bf 149}, 117 (1984);
	M. Dine, N. Seiberg and E. Witten, Nucl. Phys. B {\bf 289}, 589 (1987).

\item	R. N. Mohapatra and B. Sakita, Phys. Rev. D {\bf 21}, 1062 (1980).

\item   C. Jarlskog, Phys. Rev. D {\bf 35}, 1685 (1987); {\bf 36}
        2138 (1987); C. Jarlskog and A. Kleppe, Nucl. Phys. {\bf B286},
        245 (1987).

\item	An early observation of this effect in a Monte Carlo analysis was
	reported in C. H. Albright, Phys. Rev. D {\bf R725}, (1992).  An 
	analytical analysis was subsequently reported by A. Yu. Smirnov,
	Phys. Rev. D {\bf 48}, 3264 (1993).
\end{reflist}
\newpage
\vspace*{1in}
\begin{center}
\begin{tabular}{c|rrr|c}
	SU(5) & \multicolumn{3}{c|}{VEV Directions} & Flipped SU(5)\\
	Assignments & ${\bf 45}_X$ & ${\bf 45}_Y$ & ${\bf 45}_Z$ &
		Assignments\\ \hline
	$u,\ d$ & 1 & 1 & 1 & $d,\ u$\\
	$u^c$ & 1 & - 4 & 2 & $d^c$\\
	$d^c$ & - 3 & 2 & - 4 & $u^c$\\
	$\nu,\ \ell$ & - 3 & - 3 & - 3 & $\ell,\ \nu$\\
	$\nu^c$ & 5 & 0 & 6 & $e^c$\\
	$e^c$ & 1 & 6 & 0 & $\nu^c$\\ \hline
\end{tabular}

\vspace*{0.25in}
	Table I.  Couplings of the ${\bf 45}$ VEVs to states in the ${\bf 16}$.
\end{center}

\vspace*{1.5in}
\noindent Fig. 1.  Tree-level diagrams for the (a) renormalizable and (b) 
leading-order nonrenormalizable contributions to the generic Dirac mass matrix 
and for the (c) 33 element of the Majorana mass matrix.
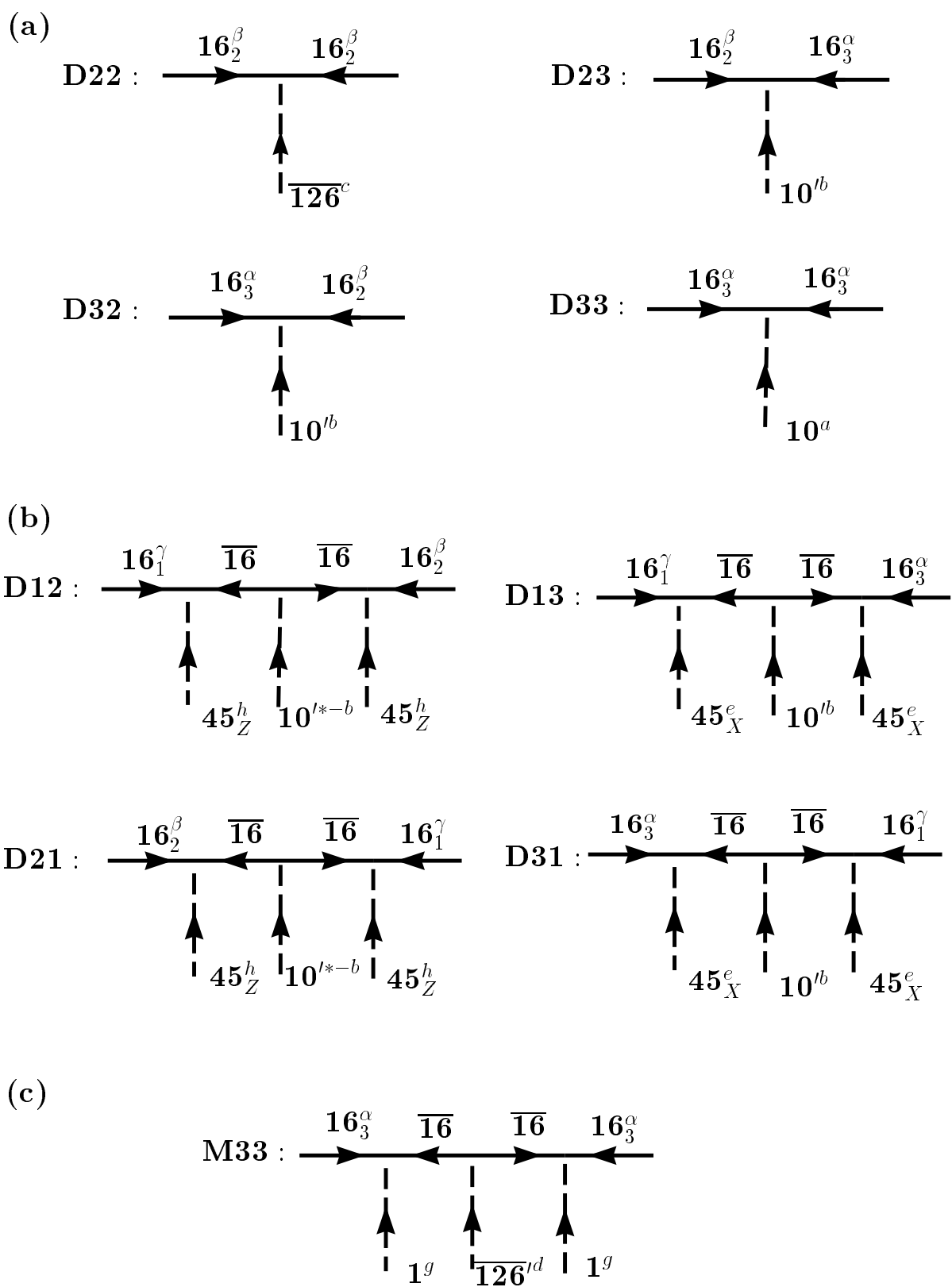
\end{document}